\documentclass[11pt]{amsart}
\usepackage{hyperref}
\def \dd{{\rm d}}
\begin{document}

\title[]{Proposal for an Interpretation of the Hermitian Theory of Relativity}
\author{S. Antoci}
\address{Dipartimento di Fisica ``A. Volta'', Via A. Bassi, 6, 27100
Pavia, Italy.}

\thanks{General Relativity and Gravitation,  {\bf 19}, 665 (1987).}

\begin{abstract}
The equilibrium conditions for charges and currents, apparent in
exact solutions of the field equations, lead one to regard the
Hermitian theory of relativity as the theory of a field endowed
with two sources: electromagnetic and colour four-currents.
\end{abstract}
\maketitle
\section{Introduction}
The Hermitian theory of relativity dates back to 1925, when
Einstein proposed it as a unitary field theory of gravitation and
electromagnetism \cite{Einstein1925}. Only in 1945, after several
attempts of a different kind, did Einstein return
\cite{Einstein1945} to the original idea, which he subsequently
pursued, together with his collaborators, in several works until
his death in 1955. Difficulties were soon encountered when trying
to find exact solutions of the field equations, which could allow
for some glimpse into the physical content of the theory
\cite{PW}, while approximate calculations led to disconcerting
results for the motion of charged particles \cite{Callaway1953}.
Both Einstein \cite{Einstein1955} and Schr\"odinger
\cite{Schroedinger1954} did not believe that the evidence brought
against the theory by the approximate calculations could be
considered really conclusive: to them, the problem of
identification of the geometrical objects of the theory with
physical entities was still an open one, since no special exact
solution, which suggested an application to anything that might
interest us, had yet been found. However, with a turn motivated
more by psychological than by rational reasons, the interest in
that theory faded, and the Hermitian theory of relativity is
considered by present-day physicists, with a few exceptions, a
subject of purely historical interest.\par
    In recent times a method was found \cite{Antoci1987},
which allows one to build solutions of the field equations of
Hermitian relativity depending on three coordinates;
particular solutions obtained by this method appear to have
direct physical meaning if singular sources, not contemplated
by the original, tentative interpretation of Einstein and
Schr\"odinger, but called in by the very structure of the
solutions, are allowed. The evidence brought by these solutions on
the merit of introducing singular sources is so compelling that a
thorough reexamination of the Hermitian extension of general
relativity, in order to fully appreciate its physical content, is
urgently required.

\section{The field equations}
By extending into the complex domain the symmetry postulates of
general relativity, let us consider the Hermitian fundamental form
$g_{ik}=g_{(ik)}+g_{[ik]}$ and the affine connection
$\Gamma^i_{kl}=\Gamma^i_{(kl)}+\Gamma^i_{[kl]}$,
Hermitian with respect to the lower indices; both entities depend
on the real coordinates $x^i$, with $i$ running from 1 to 4. We
define also the contravariant tensor $g^{ik}$ by the relation
\begin{equation}\label{1}
g^{il}g_{kl}=\delta^i_k
\end{equation}
and the contravariant tensor density
${\bf g}^{ik}=(-g)^{1/2}g^{ik}$, where $g=\text{det} (g_{ik})$.

Then the field equations of the Hermitian theory of
relativity \cite{Einstein1948} can be written as
\begin{eqnarray}\label{2}
g_{ik,l}-g_{nk}\Gamma^n_{il}-g_{in}\Gamma^n_{lk}=0,\\\label{3}
{\bf g}^{[is]}_{~~,s}=0,\\\label{4}
R_{(ik)}(\Gamma)=0,\\\label{5}
R_{[ik],l}(\Gamma)+R_{[kl],i}(\Gamma)+R_{[li],k}(\Gamma)=0;
\end{eqnarray}
$R_{ik}(\Gamma)$ is the Ricci tensor
\begin{equation}\label{6}
R_{ik}(\Gamma)=\Gamma^a_{ik,a}-\Gamma^a_{ia,k}
-\Gamma^a_{ib}\Gamma^b_{ak}+\Gamma^a_{ik}\Gamma^b_{ab}.
\end{equation}

Let $\varepsilon^{iklm}$ represent Levi Civita's tensor density;
from
$R_{[ik]}$ we can define the dual tensor density
\begin{equation}\label{7}
^*{\bf R}^{ik}=\frac12\varepsilon^{iklm}R_{[lm]},
\end{equation}
and write (\ref{5}) as
\begin{equation}\label{8}
^*{\bf R}^{[is]}_{~~,s}=0.
\end{equation}

\section{The problem of external sources}
Before looking for particular solutions of the field equations
(\ref{2})-(\ref{5}), let us remember the issue whether, in the
Hermitian theory of relativity, field singularities are to be
allowed or not.\par
    Einstein and Schr\"odinger expressed clearly their opinion on
this point: for them, singularities should not be allowed at all;
(\ref{3}), for instance, needs to be satisfied everywhere. A
comparison with Maxwell's theory then suggests that this equation
is interpreted as expressing the vanishing of the magnetic current
density; if this is accepted, the expression of the electric
current density \cite{Einstein1955} should be
${\bf J}^m=\frac16\varepsilon^{iklm}g_{[[ik],l]}$. However, no
static spherically symmetric solution that complies with such a
proposal can exist \cite{TP1970}, and the approximate solution of
the problem of motion, done according to the above-mentioned
identification, found no evidence for the Lorentz force
\cite{Callaway1953}.\par
    It was Treder who proved \cite{Treder1957} the existence of
nongravitational forces in the Hermitian theory of relativity; his
result (a Coulomb-like force plus a force independent of distance)
could only be obtained by allowing for singular sources at the
right-hand-side of (\ref{5}). Since, in Treder's approximate
calculation, the constant force cannot be made to vanish without
cancelling the Coulomb-like force itself, and since constant
forces were not considered in 1957 to be physically meaningful,
that finding seemed to provide arguments against Hermitian
relativity.\par
    In 1980 Treder himself showed \cite{Treder1980} that his
earlier finding could be given a chromodynamic interpretation.
Stimulated by this result, I propose to allow for singular sources
on the right-hand-sides of both (\ref{3}) and (\ref{5}). These
sources are conserved four-currents, and particular solutions
obtained by the method of the next section show that it is indeed
worth allowing for them, since a physical interpretation then
becomes apparent.

\section{A method for finding solutions}

    Despite the complexity of the field equations (\ref{2})-(\ref{5}),
a simple method exists \cite{Antoci1987}, that allows one to build
solutions of the Hermitian theory of relativity from known vacuum
solutions of the general theory of relativity. To quote this
result, I assume henceforth that Latin indices run
from 1 to 4, while Greek indices run from 1 to 3.\par
    Let $h_{ik}$ be the fundamental form of a vacuum
solution of general relativity, which depends only on the three
co-ordinates $x^{\lambda}$, and for which $h_{\lambda 4}=0$;
we define also a purely imaginary tensor $a_{ik}=a_{ik}(x^{\lambda})$,
and we assume that its only nonvanishing
components are $a_{\mu 4}=-a_{4 \mu}$. We form the tensor
\begin{equation}\label{9}
\alpha_i^{~k}=a_{il}h^{kl}=-\alpha^k_{~i},
\end{equation}
where $h^{ik}$ is the inverse of $h_{ik}$, and we define the
Hermitian fundamental form $g_{ik}$ as follows:
\begin{eqnarray}\nonumber
g_{\lambda\mu}=h_{\lambda\mu},\\\label{10}
g_{4\mu}=\alpha_4^{~\nu}h_{\mu\nu},\\\nonumber
g_{44}=h_{44}-\alpha_4^{~\mu}\alpha_4^{~\nu}h_{\mu\nu}.\nonumber
\end{eqnarray}

With such a choice of the fundamental tensor, (\ref{3}) reduces to
the single equation
\begin{equation}\label{11}
[(-\text{det}~h_{ik})^{1/2}\alpha_4^{~\lambda}
h^{44}]_{,\lambda}=0.
\end{equation}

In the Hermitian theory of relativity, (\ref{2}) defines the
affinity $\Gamma^i_{kl}$ in terms of the fundamental tensor; it
can be uniquely solved \cite{TH} under very general circumstances
for $g_{ik}$. We now solve (\ref{2}) for the particular choice of
the fundamental tensor given by (\ref{10}); it turns out that, if
$\alpha_i^{~k}$ obeys the three equations
\begin{equation}\label{12}
\alpha^4_{~\mu,\lambda}-\alpha^4_{~\lambda,\mu}=0,
\end{equation}
the result simplifies considerably, and the nonzero components of
$\Gamma^i_{kl}$ can be written as
\begin{eqnarray}\label{13}
\Gamma^{\lambda}_{(\mu\nu)}=\left\{^{~\lambda}_{\mu~\nu}\right\},
\\\nonumber
\Gamma^{\lambda}_{[4\nu]}=\alpha^{~\lambda}_{4~,\nu}
-\left\{^{~4}_{4~\nu}\right\}\alpha^{~\lambda}_4
+\left\{^{~\lambda}_{\rho~\nu}\right\}\alpha^{~\varrho}_4,
\\\nonumber
\Gamma^4_{(4\nu)}=\left\{^{~4}_{4~\nu}\right\},
\\\nonumber
\Gamma^{\lambda}_{44}=\left\{^{~\lambda}_{4~4}\right\}
-\alpha^{~\nu}_4\left(\Gamma^{\lambda}_{[4\nu]}
-\alpha^{~\lambda}_4\Gamma^4_{(4\nu)}\right);
\end{eqnarray}
in (\ref{13}), $\left\{^{~i}_{k~l}\right\}$ is the Christoffel
connection built with $h_{ik}$. We are now in a position to write
the Ricci tensor $R_{ik}(\Gamma)$, and we find that, if (\ref{11})
and (\ref{12}) hold, its nonvanishing components read
\begin{eqnarray}\nonumber
R_{\lambda\mu}=S_{\lambda\mu},
\\\label{14}
R_{4\mu}=\alpha^{~\nu}_4S_{\mu\nu}+\left(\alpha^{~\nu}_4
\left\{^{~4}_{4~\nu}\right\}\right)_{,\mu},
\\\nonumber
R_{44}=S_{44}-\alpha^{~\mu}_4\alpha^{~\nu}_4S_{\mu\nu},
\end{eqnarray}
where $S_{ik}$ is the Ricci tensor built with
$\left\{^{~i}_{k~l}\right\}$. But $S_{ik}=0$ for a vacuum
solution of general relativity; therefore, even the field equations (\ref{4})
and (\ref{5}) of the Hermitian theory are satisfied. Under the
circumstances mentioned above, the problem of solving (\ref{2})-(\ref{5})
reduces to a simpler task: to satisfy (\ref{11}) and (\ref{12})
for a given field $h_{ik}$, solution of general relativity. Such a
task is immediately accomplished when $h_{ik}$ corresponds to some
simple form of the Minkowski metric.

\section{Electrostatics}
    We can, for instance, start with $h_{ik}$ as representing the
Minkowski metric referred to Cartesian coordinates $x$, $y$, $z$,
$t$ and obtain the solution
\begin{equation}\label{15}
g_{ik}=\left(\begin{array}{rrrr}
 -1 &  0 &  0 & a \\
  0 & -1 &  0 & b \\
  0 &  0 & -1 & c \\
 -a & -b & -c & d
\end{array}\right),
\end{equation}
with
\begin{equation}\label{16}
d=1+a^2+b^2+c^2
\end{equation}
and
\begin{equation}\label{17}
a=i\chi_{,x}, \ b=i\chi_{,y}, \ c=i\chi_{,z}, \ \
\chi_{,xx}+\chi_{,yy}+\chi_{,zz}=0.
\end{equation}

The imaginary part $g_{[ik]}$ of this solution just looks like the
general electrostatic solution of Maxwell's theory. If we allow
for singularities in (\ref{3}), the electrostatic field due to $n$
point charges $h_q$, located at $x=x_q$, $y=y_q$, $z=z_q$, can be
built by taking
\begin{equation}\label{18}
\chi=-\sum_{q=1}^n\frac{h_q}{p_q},
\end{equation}
where
\begin{equation}\label{19}
p_q=[(x-x_q)^2+(y-y_q)^2+(z-z_q)^2]^{1/2}.
\end{equation}

But a nonlinear theory like ours should give more information than
Max\-well's equations: it should predict also the equations of
motion, i.e., in this case, the conditions for the electrostatic
equilibrium of charges $h_q$. We have agreed to represent charges
by singularities, and previous experience with the problem of
motion in general relativity has shown \cite{EI1949} that the
shape of the singularities needs to be restricted in order to get
a meaningful result. We can learn from the solution what kind of
restriction must be imposed; we expect that such a restriction may
arise from terms which are nonlinear in charges $h_q$. Let us
consider the only nonlinear term which appears in $g_{ik}$:
\begin{eqnarray}\label{20}
d=1-\sum^n_{q=1}\frac{h_q^2}{p_q^4}\\\nonumber
-\sum^{n (q\neq q')}_{q,q'=1}
h_qh_{q'}\frac{(x-x_q)(x-x_{q'})+(y-y_q)(y-y_{q'})
+(z-z_q)(z-z_{q'})}{p_q^3p_{q'}^3}.\nonumber
\end{eqnarray}

If we want to avoid that $d$ must exhibit a singularity in the form of
a jump, arising from the presence of the other charges, at the
position where the $q$th charge is located, we have to require
that
\begin{equation}\label{21}
\sum^n_{q'\neq q}h_{q'}\frac{x_q-x_{q'}}{r_{qq'}^3}
=\sum^n_{q'\neq q}h_{q'}\frac{y_q-y_{q'}}{r_{qq'}^3}
=\sum^n_{q'\neq q}h_{q'}\frac{z_q-z_{q'}}{r_{qq'}^3}
=0,
\end{equation}
where $r_{qq'}$ is the Cartesian distance between the two charges
$q$ and $q'$. Equation (\ref{21}) indeed expresses the equilibrium
conditions for $n$ point electric charges; when $g_{(ik)}$ is assumed
to be the metric tensor of the Hermitian theory, this result
is true not only in the coordinate space, but also in the physical space;
the no-jump rule appears to be a physically relevant restriction
on singularities.

\section{Electric currents in equilibrium}

When $h_{ik}$ is given by the Minkowski metric referred to
Cartesian coordinates, we can get also another solution, which
depends on $x$, $y$ and $t$:
\begin{equation}\label{22}
g_{ik}=\left(\begin{array}{rrrr}
 -1 &  0 &  e & 0 \\
  0 & -1 &  f & 0 \\
 -e & -f &  h & c \\
  0 &  0 & -c & 1
\end{array}\right),
\end{equation}
with
\begin{equation}\label{23}
h=-1+e^2+f^2-c^2
\end{equation}
and
\begin{equation}\label{24}
e=i\xi_{,x}, \ f=i\xi_{,y}, \ c=-i\xi_{,t}, \ \
\xi_{,xx}+\xi_{,yy}-\xi_{,tt}=0;
\end{equation}
even in this case $g_{[ik]}$ obeys Maxwell's equations.\par
We allow for singularities in (\ref{3}) and consider the
particular solution for which
\begin{equation}\label{25}
\xi=\sum_{q=1}^n l_q\ln p_q,
\end{equation}
where
\begin{equation}\label{26}
p_q=[(x-x_q)^2+(y-y_q)^2]^{1/2}.
\end{equation}

In this solution $g_{[ik]}$ represents the magnetic field generated by $n$
steady currents; each current, with intensity $l_q$, is
running along a wire parallel to the $z$ axis, which intersects
the $x$, $y$ plane at $x=x_q$, $y=y_q$. The no-jump rule, which
now restricts the nonlinear term $h$, requires that
\begin{equation}\label{27}
\sum^n_{q'\neq q}l_{q'}\frac{x_q-x_{q'}}{r_{qq'}^2}
=\sum^n_{q'\neq q}l_{q'}\frac{y_q-y_{q'}}{r_{qq'}^2}
=0,
\end{equation}
where $r_{qq'}$ is the Cartesian distance between the two wires
$q$ and $q'$. Equation (\ref{27}), as expected, predicts the
equilibrium positions of $n$ parallel wires at rest, run by steady
currents.\par
    As a further example, we can take
\begin{eqnarray}\label{28}
\xi=\sum_{q=1}^n\frac{l_q}{p_q},\\\nonumber
p_q=[(t-t_q)^2-(x-x_q)^2-(y-y_q)^2]^{1/2}.
\end{eqnarray}
In the sum each term is considered only if
\begin{equation}\label{29}
(t-t_q)^2>(x-x_q)^2+(y-y_q)^2.
\end{equation}

In such a solution $g_{[ik]}$ represents the superposition of
cylindrical electromagnetic waves radiated by $n$ current pulses,
occurring in wires parallel to the $z$ axis; the $q$th pulse
occurs at time $t=t_q$ on the wire which intersects the $x$, $y$
plane at $x=x_q$, $y=y_q$; both advanced and retarded fields with
equal weight are taken. The no-jump condition, when applied to the
singularities of this solution, prescribes that
\begin{equation}\label{30}
\sum^n_{q'\neq q}l_{q'}\frac{t_q-t_{q'}}{s_{qq'}^3}
=\sum^n_{q'\neq q}l_{q'}\frac{x_q-x_{q'}}{s_{qq'}^3}
=\sum^n_{q'\neq q}l_{q'}\frac{y_q-y_{q'}}{s_{qq'}^3}
=0,
\end{equation}
where
\begin{equation}\label{31}
s_{qq'}=[(t_q-t_{q'})^2-(x_q-x_{q'})^2-(y_q-y_{q'})^2]^{1/2}.
\end{equation}

Again, a physically correct equilibrium condition, in agreement
with electromagnetism, is found. When (\ref{30}) holds, the
electromagnetic field due to the other current pulses vanishes at
the time and place where the $q$th current pulse is occurring; no
allowance is made for exchanges of energy and momentum between the
current pulses.\par

\section{Electric charges in uniformly accelerated motion}

    It would be interesting to provide an example of a solution
allowing for exchanges of energy and momentum which were absent
in the solutions considered up to now; as everybody knows, such
exchanges occur when an electric charge is accelerating under the
influence of an electromagnetic field. One such example can be
found if we take for $h_{ik}$ the form
\begin{equation}\label{32}
h_{ik}=\left(\begin{array}{rrrr}
 -1 &  0 &  0 & 0 \\
  0 & -1 &  0 & 0 \\
  0 &  0 & -1 & 0 \\
  0 &  0 &  0 & z^2
\end{array}\right),
\end{equation}
referred to Cartesian coordinates $x$, $y$, $z$, $t$, and we build
the solution
\begin{equation}\label{33}
g_{ik}=\left(\begin{array}{rrrr}
 -1 &  0 &  0 & a \\
  0 & -1 &  0 & b \\
  0 &  0 & -1 & c \\
 -a & -b & -c & d
\end{array}\right),
\end{equation}
with
\begin{equation}\label{34}
d=z^2+a^2+b^2+c^2
\end{equation}
and
\begin{equation}\label{35}
a=iz^2\chi_{,x}, \ b=iz^2\chi_{,y}, \ c=iz^2\chi_{,z}, \ \
\chi_{,xx}+\chi_{,yy}+\chi_{,zz}+(\chi_{,z}/z)=0.
\end{equation}

We allow for singularities in (\ref{3}) and consider the
particular solution with
\begin{equation}\label{36}
\chi=-\sum_{q=1}^n\frac{K_q}{p_q}+m\ln z,
\end{equation}
where
\begin{eqnarray}\label{37}
p_q=[(s+h_q)^2-2h_q z^2]^{1/2},\\\nonumber
s=\frac12(x^2+y^2+z^2);
\end{eqnarray}
$K_q$, $h_q$ and $m$ are constants; we assume $h_q>0$. We get
\begin{eqnarray}\label{38}
a=i\sum_{q=1}^n\frac{K_qz^2x(s+h_q)}{p_q^3},\\\nonumber
b=i\sum_{q=1}^n\frac{K_qz^2y(s+h_q)}{p_q^3},\\\nonumber
c=i\sum_{q=1}^n\frac{K_qz^3(s-h_q)}{p_q^3}+imz,
\end{eqnarray}
and
\begin{equation}\label{39}
d=z^2(1-F),
\end{equation}
where
\begin{eqnarray}\nonumber
F=2sz^2\sum_{q=1}^n\frac{K_q^2}{p_q^4}+m^2+z^2(x^2+y^2)
\sum_{q,q'=1}^{n (q\ne q')}
\frac{K_qK_{q'}(s+h_q)(s+h_{q'})}{p_q^3p_{q'}^3}\\\label{40}
+z^4\sum_{q=1}^n\frac{K_q(s-h_q)}{p_q^3}
\left(\frac{2m}{z^2}
+\sum_{q'\ne q}^n\frac{K_{q'}(s-h_{q'})}{p_{q'}^3}\right).
\end{eqnarray}

To understand the meaning of this solution, let us perform the
coordinate transformation
\begin{equation}\label{41}
r'=(x^2+y^2)^{1/2}, \ z'=z\cosh t, \
\varphi'=\arctan(y/x), \ t'=z\sinh t,
\end{equation}
and look at the transformed tensor $g'_{ik}$. Its nonzero
components are:
\begin{eqnarray}\label{42}
g'_{11}=-1,\\\nonumber
g'_{21}=i\sum_{q=1}^n\frac{K_qr't'(s'+h_q)}{{p'}_q^3},\\\nonumber
g'_{41}=-i\sum_{q=1}^n\frac{K_qr'z'(s'+h_q)}{{p'}_q^3},\\\nonumber
g'_{22}=-1-\frac{t'^2}{z'^2-t'^2}F',\\\nonumber
g'_{42}=\frac{z't'}{z'^2-t'^2}F'
-i\sum_{q=1}^n\frac{K_q(z'^2-t'^2)(s'-h_q)}{{p'}_q^3}
-im,\\\nonumber
g'_{33}=-r'^2,\\\nonumber
g'_{44}=1-\frac{z'^2}{z'^2-t'^2}F'.
\end{eqnarray}
In (\ref{42}), $s'$, $p'_q$, $F'$ are those functions of the
primed variables $r'$, $z'$, $t'$ which are obtained from
$s$, $p_q$, $F$ when the change of variables is performed in
them.\par
The form (\ref{42}) does not hold as a solution only when
$z'^2>t'^2$, as it would appear from the coordinate transformation
given by (\ref{41}); it is valid in the whole primed space-time.
This fact can be readily ascertained by building from the metric
\begin{equation}\label{43}
h_{ik}=\left(\begin{array}{rrrr}
 -1 &  0 &  0 & 0 \\
  0 & -1 &  0 & 0 \\
  0 &  0 & -t^2 & 0 \\
  0 &  0 &  0 & 1
\end{array}\right),
\end{equation}
referred to Cartesian coordinates $x$, $y$, $z$, $t$, a Hermitian
solution which, under the coordinate transformation
\begin{equation}\label{44}
r'=(x^2+y^2)^{1/2}, \ z'=t\sinh z, \
\varphi'=\arctan(y/x), \ t'=t\cosh z,
\end{equation}
reduces to the form (\ref{42}) for $z'^2<t'^2$; the two halves
join smoothly along the hypersurfaces $z'^2=t'^2$.\par
The solution displays essential singularities only at the points
for which
\begin{equation}\label{45}
r'=0, \ z'^2-t'^2=2h_q, \ q=1,...,n.
\end{equation}
Let us evaluate the integral
\begin{equation}\label{46}
I=\frac{1}{8\pi i}\int_{\Sigma}{{\bf g}'}^{[ik]}
\dd f^*_{ik}
\end{equation}
where $\dd f^*_{ik}$ is the surface element of an ordinary surface
$\Sigma$ which envelops one of the singularities; we find $I=K_q$.
We can interpret this solution as describing the field of $2n$
point electric charges $K_q$ which perform accelerated motions
along the $z'$ axis in the presence of a constant electric field,
with intensity $m$, directed along the same axis. Each charge
performs a hyperbolic motion, with acceleration
$a=\pm(2h_q)^{-1/2}$, with respect to the coordinate space; in the
limit, when both the charges and the external field tend to
vanishing values, the motion results to be hyperbolic even in the
physical space. The motion of $n$ of these charges occurs entirely
in the half-space with $z'>0$, and it is the mirror image of the
motion of the other $n$ charges, with respect to the plane
$z'=0$.\par
We can restrict the analysis to what goes on in the half-space
with $z'>0$; in fact, due to their peculiar motion, the charges
which are in the half-space with $z'>0$ do not interact with the
charges which are in the other half-space. According to the
no-jump rule, which in this case restricts the nonlinear term
$F'$, only those motions are allowed for which
\begin{equation}\label{47}
K_q m+\sum_{q'\ne q}^n
\frac{2K_qK_{q'}h_{q'}(h_{q}-h_{q'})}{|h_{q}-h_{q'}|^3}
+\sum_{q'\ne q}^n\frac{2K_qK_{q'}}{|h_{q}-h_{q'}|}=0.
\end{equation}

This regularity condition can be interpreted physically as an
equilibrium condition for the forces acting ot the $q$th charge.
We recognize, in the first term of (\ref{47}), the force exerted
by the constant electric field with intensity $m$. The second term
is the sum of the electric forces exerted by the charges $K_{q'}$
in hyperbolic motion \cite{Born1909} on the charge $K_q$. The last
term is new, and I interpret it as expressing the inertial force
felt by the charge $K_q$. The existence of such a term is expected
on physical grounds, since the charge $K_q$ must acquire inertia
due to its very interaction with the other charges. Obviously,
this interpretation needs to be confirmed by further examples; it
would be a nice feature of the Hermitian theory if in it the
existence of masses could be derived from the existence of
charges, rather than postulated independently as occurs in general
relativity.

\section{Colour sources}

Up to now, only solutions with singular sources at the right-hand
side of (\ref{3}) have been considered; they can be named
electromagnetic. But we can also allow for singular sources in
(\ref{8}), the equivalent of field equation (\ref{5}). Both
(\ref{3}) and (\ref{8}), in fact, allow for the addition of a
conserved four-current at the right-hand side.\par
We can find a solution with this behaviour if we start from the
Minkowski metric $h_{ik}$ referred to polar cylindrical
coordinates $r$, $z$, $\varphi$, $t$:
\begin{equation}\label{48}
h_{ik}=\left(\begin{array}{rrrr}
 -1 &  0 &  0 & 0 \\
  0 & -1 &  0 & 0 \\
  0 &  0 & -r^2 & 0 \\
  0 &  0 &  0 & 1
\end{array}\right).
\end{equation}
We get the solution \cite{Antoci1984}
\begin{equation}\label{49}
g_{ik}=\left(\begin{array}{rrrr}
  -1 & 0 & \delta & 0 \\
  0 & -1 & \varepsilon & 0 \\
  -\delta & -\varepsilon & \zeta & \tau \\
  0 & 0 & -\tau & 1
\end{array}\right),
\end{equation}
with
\begin{equation}\label{50}
\zeta=-r^2+\delta^2+\varepsilon^2-\tau^2,
\end{equation}
and
\begin{equation}\label{51}
\delta=ir^2\psi_{,r}, \ \varepsilon=ir^2\psi_{,z},
\ \tau=-ir^2\psi_{,t}, \ \
\psi_{,rr}+(\psi_{,r}/r)+\psi_{,zz}-\psi_{,tt}=0.
\end{equation}

Let us consider the particular solution for which
\begin{equation}\label{52}
\psi=-\sum_{q=1}^n K_q\ln\frac{p_q+z-z_q}{r}+m\ln r,
\end{equation}
where
\begin{equation}\label{53}
p_q=[r^2+(z-z_q)^2]^{1/2};
\end{equation}
$K_q$, $z_q$ and $m$ are constants. We get
\begin{eqnarray}\label{54}
\delta=i\sum^n_{q=1}\frac{K_q r(z-z_q)}{p_q}+imr,\\\nonumber
\varepsilon=-i\sum^n_{q=1}\frac{K_q r^2}{p_q}, \ \tau=0,
\end{eqnarray}
and
\begin{equation}\label{55}
\zeta=-r^2(1+F),
\end{equation}
where
\begin{eqnarray}\label{56}
F=\sum^n_{q=1}K_q^2+m^2
+r^2\sum^{n (q\neq q')}_{q,q'=1}
\frac{K_qK_{q'}}{p_qp_{q'}}\\\nonumber
+\sum^n_{q=1}\frac{K_q(z-z_q)}{p_q}
\left[2m+\sum^n_{q'\ne q}\frac{K_{q'}(z-z_{q'})}{p_{q'}}\right].
\end{eqnarray}
We go over to Cartesian coordinates with the transformation
\begin{equation}\label{57}
x'=r\cos\varphi, \ y'=r\sin\varphi, \ z'=z, \ t'=t;
\end{equation}
the nonzero components of the transformed tensor $g'_{ik}$ read
\begin{eqnarray}\label{58}
g'_{11}=-1-\frac{y'^2}{x'^2+y'^2}F',\\\nonumber
g'_{21}=\frac{x'y'}{x'^2+y'^2}F'
-i\sum_{q=1}^n\frac{K_q(z'-z_q)}{{p'}_q}
-im,\\\nonumber
g'_{31}=i\sum_{q=1}^n\frac{K_qy'}{{p'}_q},\\\nonumber
g'_{22}=-1-\frac{x'^2}{x'^2+y'^2}F',\\\nonumber
g'_{32}=-i\sum_{q=1}^n\frac{K_qx'}{{p'}_q},\\\nonumber
g'_{33}=-1,\\\nonumber
g'_{44}=1.
\end{eqnarray}
$p'_q$ and $F'$ are those functions of $x'$, $y'$, $z'$ which can
be obtained from $p_q$ and $F$ through the change of variables.
The solution displays essential singularities at
\begin{equation}\label{59}
x'=y'=0, \ z'=z_q, \ q=1,...,n.
\end{equation}
Let us evaluate the integral
\begin{equation}\label{60}
I=-\frac{1}{8\pi i}\int_{\Sigma}R_{[ik]}
\dd f^{ik},
\end{equation}
extended to an ordinary surface $\Sigma$ which envelops one
of the singularities. We get $I=K_q$. It seems therefore that we
have to do with $n$ charges of a new kind, at rest on the $z'$
axis, in the presence of a constant magnetic field of intensity
$m$, directed along the same axis. The no-jump rule for
singularities yields the equilibrium condition
\begin{equation}\label{61}
\sum_{q'\ne q}^n\frac{K_qK_{q'}(z_q-z_{q'})}{|z_q-z_{q'}|}
+K_qm=0,
\end{equation}
which we interpret as expressing a force balance. The first term
of (\ref{61}) shows that the new charges interact mutually with
forces which do not depend on the distance between them; this
occurs in the coordinate space and even in the physical space if
$g_{(ik)}$ is the metric tensor. As was mentioned in Section 3,
such forces were found in 1957 by Treder with an approximate
calculation \cite{Treder1957} and were recently interpreted
\cite{Treder1980} by the same author as confining forces between
colour charges. The second term of (\ref{61}) shows that the
colour charges predicted by the Hermitian theory of relativity,
when sitting at rest in a constant magnetic field, feel a constant
pull, as magnetic monopoles would do.\par
This result does not mean that the new charges are magnetic
monopoles; in fact, the field $g_{[ik]}$ does not obey Maxwell's
equations, and we have just shown that the charges $K_q$ cannot
exist in a free state, since they are bound by constant forces. A
mixed behaviour of this kind is at variance with the predictions
of quantum chromodynamics, but we cannot conclude from here that
it does not agree with experience. Of course, such new behaviour
can be expected from a truly unified theory, in which all the
fields merge in a single entity \cite{Schroedinger1954} without
artificial boundaries between them.

\section{Conclusion}

If allowance is made for singular, conserved sources at the
right-hand sides of (\ref{3}) and (\ref{8}), then particular exact
solutions to the field equations of the Hermitian theory of
relativity find a direct physical interpretation. These solutions
show that past negative results found with this theory cannot be
imputed to the field equations per se, but only to a false
identification of the physical entities with the geometrical
objects of the theory. With a new identification, the outcome of
Einstein's lifelong effort to achieve a unified field theory no
longer appears to be a failure, as is currently believed. It is
indeed worth investigating again how far the Hermitian theory of
relativity can go in accounting for gravitational, electromagnetic
and nuclear phenomena.
\bigskip

{\bf Aknowledgments.} I am indebted to Professor E. Kreisel and Professor
H.-J. Treder for many helpful discussions and for the support given to
my work.

\bibliographystyle{amsplain}

\end{document}